\documentclass{emulateapj}
\usepackage{graphicx}
\usepackage{multirow,natbib}
\usepackage{textcomp}
\usepackage{epsfig}
\usepackage{amsmath}
\usepackage{amssymb}
\usepackage{amsthm}
\usepackage{xy}

\begin{document}

\title{An Extraordinary Outburst of the Magnetar Swift J1822.3-1606}
\author{Manoneeta Chakraborty$^{1}$, Ersin G\"o\u{g}\"u\c{s}$^{1}$}
\affil{$^{1}$ Sabanc\i~University, Faculty of Engineering and Natural
  Sciences, Orhanl\i ~Tuzla 34956 \.Istanbul Turkey}

\begin{abstract}\label{Abstract}
The 2011 outburst of Swift J1822.3--1606 was extraordinary; periodic modulations at the spin period of the underlying neutron star were clearly visible, remarkably similar to what is observed during the decaying tail of magnetar giant flares. We investigated the temporal characteristics of X-ray emission during the early phases of the outburst. We performed a periodicity search with the spectral hardness ratio (HR), and found a coherent signal near the spin period of the neutron star, but with a lag of about 3 radians. Therefore, the HR is strongly anti-correlated with the X-ray intensity, which is also seen in the giant flares. We studied time evolution of the pulse profile and found that it evolves from a complex morphology to a much simpler shape within about a month. Pulse profile simplification also takes place during the giant flares, but on a much shorter timescale of about few minutes. We found that the amount of energy emitted during the first 25 days of the outburst is comparable to what was detected in minutes during the decaying tail of giant flares. Based on these similarities, we suggest that the triggering mechanisms of the giant flares and the magnetar outbursts are likely the same. We propose that the trapped fireball that develops in the magnetosphere at the onset of the outburst radiates away efficiently in minutes in magnetars exhibiting giant flares, while in other magnetars, such as Swift J1822.3--1606, the efficiency of radiation of the fireball is not as high and, therefore, lasts much longer. 
\end{abstract}


\section{Introduction}\label{Introduction}

Magnetars are highly magnetized ($10^{14}-10^{15}$ G) isolated neutron stars \citep{Duncan1992} with spin periods in the range 2-12 s and persistent X-ray emission of $10^{33}-10^{36}$ erg s$^{-1}$. Magnetars are usually sub-divided into two classes: anomalous X-ray pulsars (AXP) and soft gamma repeaters (SGR; see \citet{Mereghetti2015} for a recent review). Magnetars occasionally emit energetic bursts in X$-$/soft $\gamma$-rays, which last from a small fraction of a second to minutes. These bursts release energy in a wide range of $10^{38}-10^{47}$ erg s$^{-1}$ and are sometimes observed along with a long-lived outburst lasting months to years. The most energetic bursts from magnetars, the giant flares, are distinct from short magnetar bursts in many respects.   

Giant flares were observed only from three magnetars: SGR 0526-66 \citep{Mazets1979, Cline1980}, SGR 1900+14 \citep{Feroci1999, Hurley1999}, and SGR 1806-20 \citep{Hurley2005, Mereghetti2005, Palmer2005}. They lasted for about a few minutes, with peak fluxes reaching up to 10$^{46}$ erg s$^{-1}$, which is $5-6$ orders of magnitude higher than typical magnetar bursts. These events are usually characterized by a spectrally hard initial spike, lasting $\lesssim$ a second, followed by a few-hundreds-of-seconds-long tail. The periodic modulations at the spin period of the underlying magnetar could clearly be observed during the tail of the giant flares \citep{Mazets1979, Hurley1999, Mazets1999}. 

Giant flares are expected to occur as a result of sudden magnetic field reconfiguration, which could lead to the fracturing of solid neutron star crust at global scale \citep{Thompson1995, Thompson2001, Woods2001}. \citet{Feroci1999} reported that the pulse profile during the August 27 giant flare of SGR 1900+14 displayed a complex structure (implying complex magnetic field geometry) that rapidly evolved with time \citep{Woods2001}. The pulse profile contained sub-pulses that appeared and disappeared as the outburst progressed. The phases of those sub-pulses usually showed a stable behavior with time. From the spectral point of view, the modulating tail of the giant flares exhibits remarkable variations. Using BeppoSAX data, \citet{Feroci1999} found a ``see--saw" behavior in spectral hardness ratio (HR), which strongly anti-correlated with the intensity. \citet{Mazets1999} also obtained the complex four peaked pulse profile and the anti-correlation of HR and intensity for the August 27 giant flare using Konus-Wind data. A double exponential model was invoked to describe the decay of the intensity envelope of the giant flare of SGR 1900+14 \citep{Feroci1999}. The dramatic changes in the pulse profile of SGR 1900+14 during the tail of its giant flare, as well as the observed anti-correlation between hardness and intensity were interpreted with a trapped fireball in the magnetosphere of a magnetar \citep{Feroci2001}. In particular, the occurrence of the maximum of the hardness at the pulse minimum was attributed to the Comptonization by Alfv\'en waves in the extended corona or to the acceleration of non-thermal particles \citep{Thompson2001}. Excitation of toroidal modes, possibly by the global scale fracturing of the neutron star crust, has been accredited to produce the observed quasi-periodic oscillations during the tail of giant flare from SGR 1900+14 \citet{Strohmayer2005}, as well as the tail of the 2004 December 27 giant flare from SGR 1806--20 \citet{Israel2005}.

On 2014 July 14, the Burst Alert Telescope (BAT) on board \textit{Swift} detected SGR-like hard X-ray bursts, which were identified as having come from a new source,  Swift J1822.3--1606 \citep{Cummings2011}. The spin period of this source was determined to be 8.4377 s \citep{Gogus2011}. This source was classified as a magnetar on the basis of its spectral and timing properties and was of immediate interest as a new addition to the class of relatively low inferred dipole magnetic field SGRs \citep{Livingstone2011, Rea2012, Scholz2012, Scholz2014}. \citet{Rea2012} reported the values of the spin period and period derivative as $8.43772016(2)$ s and $8.3(2)\times10^{-14}$ s s$^{-1}$, respectively. The inferred dipolar magnetic field strength is $2.7 \times 10^{13}$ G, which is slightly below the magnetic field strength corresponding to electron cyclotron resonance, i.e.,  $4.4 \times 10^{13}$ G.
Such a low inferred $B$ value has been previously observed in the case of magnetar SGR 0418+5729 \citep{Rea2010,Rea2012}, where the magnetic field strength was measured to be $6 \times 10^{12}$ G. The energetic bursts which led to the discovery of Swift J1822.3--1606 marked the onset of a long lasting outburst of its persistent X-ray emission. \citet{Livingstone2011} investigated the variation of the total and pulsed X-ray flux throughout the outburst using data collected with \textit{RXTE}, \textit{Swift} and \textit{Chandra}. They were able to describe the temporal evolution of its persistent X-ray intensity using double and single exponential models. Remarkably, the periodic modulations exactly at the spin period were visible in the \textit{RXTE} observations during the early phases of its outburst.

In this paper, we investigated the observational features of the 2011 outburst of Swift J1822.3--1606 using data collected with Rossi X-ray Timing Explorer  (\textit{RXTE}), and compared them to those observed in the decaying tail of giant flares. Our aim is to examine the hypothesis of whether the long lasting 2011 outburst of Swift J1822.3--1606 resembles a somewhat weak, repressed giant flare. We present the data used, our data analysis details, and the results of our temporal and spectral examinations in the next section. We discuss the implications of our results in \S~\ref{Discussion}.

\section{Observations, Data Analysis And Results}\label{DataAnalysis}

\textit{RXTE} monitored the 2011 outburst of the source, Swift J1822.3--1606 from July 16 to November 20 with 60 pointed observations. These corresponded to a total exposure time of $177$ ks. We used the data collected with the Proportional Counter Array (PCA) on board \textit{RXTE}, which consisted of 5 xenon filled proportional counter units (PCU) operated in the energy range of 2--60 keV \citep{Jahoda2006}. It had a field of view of $1^\circ \times 1^\circ$ and an excellent time resolution of $1 \mu$s in GoodXenon mode. For our timing analysis, we employed the data of this mode to benefit from its invaluable temporal and spectral (256 channels) capabilities. For each pointing, we first screened the light curve and excluded the times of short magnetar bursts, and for the pulse timing analysis, all event arrival times were converted to that of the solar system barycenter. For our investigation, we concentrated on the first 25 days, starting from 2011 July 15, where the pulsations could clearly be sighted in the light curve. This corresponded to 25 pointings on the source totaling an effective observation time of $73$ ks. The starting times of these 25 observations are listed in Table~\ref{hrlag}. In the following sections, we describe in detail the techniques employed and the consequent results obtained with the aim of explaining the observational characteristics of the 2011 outburst of Swift J1822.3--1606.

\subsection{Variation of HR}

\begin{figure}[H!tba]
\centering
\includegraphics[width=8cm]{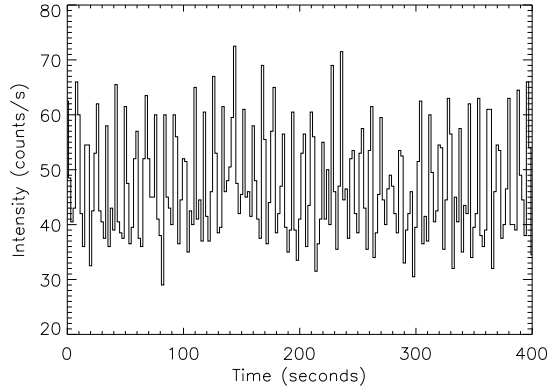}
\caption{Light curve of observation on 2011 July 19 (Observation ID: $96048-02-01-01$) during the initial stages of the tail of the outburst. Clear periodic intensity modulations at the spin period of the magnetar can be observed. \label{lc}}
\end{figure}

We first investigated whether there exists a correlative behavior of spectral hardness with intensity during the early episodes of the outburst, during which pulsed modulations could be clearly seen in the light curve, as was observed during the decaying tail of the SGR 1900+14 giant flare \citep{Feroci2001}. To demonstrate these clear pulses of X-ray emission, we present a segment of PCA light curve in the energy range of $2-20$ keV with 2 s time resolution PCA observations in Figure~\ref{lc}.

\begin{figure}[H!tba]
\centering
\includegraphics[width=8cm]{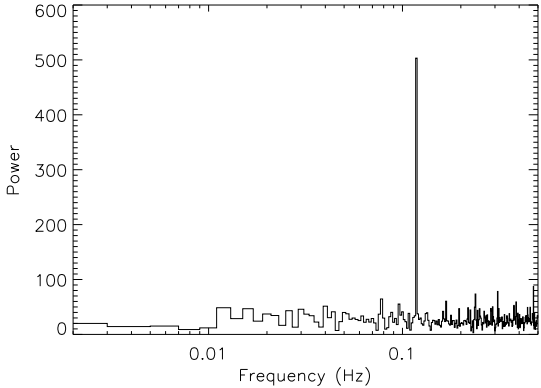}
\caption{Power spectrum of the HR for the observation on 2011 July 16. Highly significant signal power can be observed from this power spectrum at the pulsar frequency 0.118 Hz, which implies strong periodic modulation of the HR due to the pulsation. \label{hrpwspc}}
\end{figure}

We defined the HR as the ratio of the source intensity in the energy range 5.0--14.0 to that in the energy range 2.0--5.0 keV. The HR was calculated at each 1 s time step for all barycenter-corrected photon arrival times. We then proceeded to search for any periodic behavior in the HR time series. To investigate this, we performed a fast Fourier transform (FFT) of each HR series. We then computed the corresponding power spectrum of the HRs over each 500 s segment of an observation, and averaged over them to get the mean power spectrum. The resulting power spectra had a frequency resolution of 0.002 Hz and a Nyquist frequency of 0.5 Hz. The power spectra of the HR showed the most prominent peaks at the pulsar's spin frequency, indicating that the hardness becomes strongly modulated at the pulsar frequency. In Figure~\ref{hrpwspc}, we show an example of the presence of a highly significant peak in the power spectrum of the HR for the observation performed on 2011 July 16.

Next, the important issue to address is whether the source intensity and its HR exhibit a coherent phenomenon or if there exists a lag between the two. To check this, we also calculated the lag between these two time series in the following manner. Along with the FFT of the HR, we computed the FFT of the total intensity in the 2.0--14 keV energy range in each 500 s time segment of each observation. We then calculated cross-spectrum from the two FFTs in each time segment and obtained the averaged cross-spectrum. From the averaged cross-spectrum, we calculated the phase lag between the HR and the total intensity at the peak power frequency. The frequency corresponding to the maximum power was obtained from the averaged HR power spectrum of each observation and in all cases it was found to be 0.118 Hz, which is the pulsar frequency at the chosen frequency resolution. In Table.~\ref{hrlag}, we list the phase lags between the hardness and the pulse profile intensity for all the observations at 0.118 Hz during the first 25 days of \textit{RXTE} observation of the outburst. The phase lag between the hardness and the intensity is primarily negative implying that the hardness follows the intensity. The value of the lag is quite high, its magnitude always being greater than 2.5 radian, making the lag usually close to $-\pi$, which suggests a strong anti-correlation between the two quantities. It is interesting to note that there exists a marginal change in phase lag: soon after the onset it is about -2.6 rad, while later on it is around $-2.8$ to $-3$. Also note that in one case the value of the lag is positive, likely due to the fact that the pulsed signal strength was much weaker about 20 days after the outburst onset, which may have lead to the incorrect estimation of the sign of the lag.

\begin{table}[H!tba]
\centering
\caption{Table of phase lag between hardness and intensity at peak power frequency for each observation during the first 25 days of the 2011 outburst of Swift J1822.3-1606. \label{hrlag}}
\begin{tabular}{lll}
\hline
Date & Time  & Phase Lag \\
     & (UTC) &  (rad) \\
\hline
2011 Jul 16 & 10:20:00  & $-2.55\pm0.07$\\
2011 Jul 16 & 11:51:28  & $-2.56\pm0.07$\\
2011 Jul 19 & 12:25:46  & $-2.58\pm0.05$\\
2011 Jul 19 & 13:29:20  & $-2.74\pm0.04$\\
2011 Jul 20 & 10:22:30  & $-2.80\pm0.07$\\
2011 Jul 20 & 11:26:24  & $-2.83\pm0.04$\\
2011 Jul 21 & 14:04:50  & $-2.73\pm0.10$\\
2011 Jul 21 & 15:50:24  & $-2.87\pm0.19$\\
2011 Jul 21 & 09:24:12  & $-2.81\pm0.04$\\
2011 Jul 18 & 19:10:24  & $-2.77\pm0.03$\\
2011 Jul 22 & 13:36:00  & $-2.91\pm0.06$\\
2011 Jul 22 & 15:19:28  & $-2.90\pm0.09$\\
2011 Jul 23 & 09:57:26  & $-2.89\pm0.03$\\
2011 Jul 23 & 11:30:24  & $-2.90\pm0.08$\\
2011 Jul 25 & 13:45:54  & $-2.78\pm0.02$\\
2011 Jul 27 & 07:56:20  & $-2.78\pm0.05$\\
2011 Jul 29 & 06:57:22  & $-3.03\pm0.07$\\
2011 Jul 29 & 08:30:24  & $-2.81\pm0.06$\\
2011 Aug 01 & 07:01:34  & $-2.92\pm0.04$\\
2011 Aug 01 & 08:34:24  & $-2.93\pm0.12$\\
2011 Aug 04 & 20:01:28  & $-2.98\pm0.10$\\
2011 Aug 04 & 21:39:24  & $ 3.10\pm0.03$\\
2011 Aug 07 & 10:18:46  & $-3.03\pm<0.01$\\
2011 Aug 07 & 11:56:32  & $-3.08\pm0.08$\\
2011 Aug 09 & 12:31:28  & $-2.88\pm0.10$\\
\hline
\end{tabular}
\end{table}

\begin{figure}[H!tba]
\centering
\begin{tabular}{c}
\vspace{-0.8cm}
\includegraphics[width=8cm]{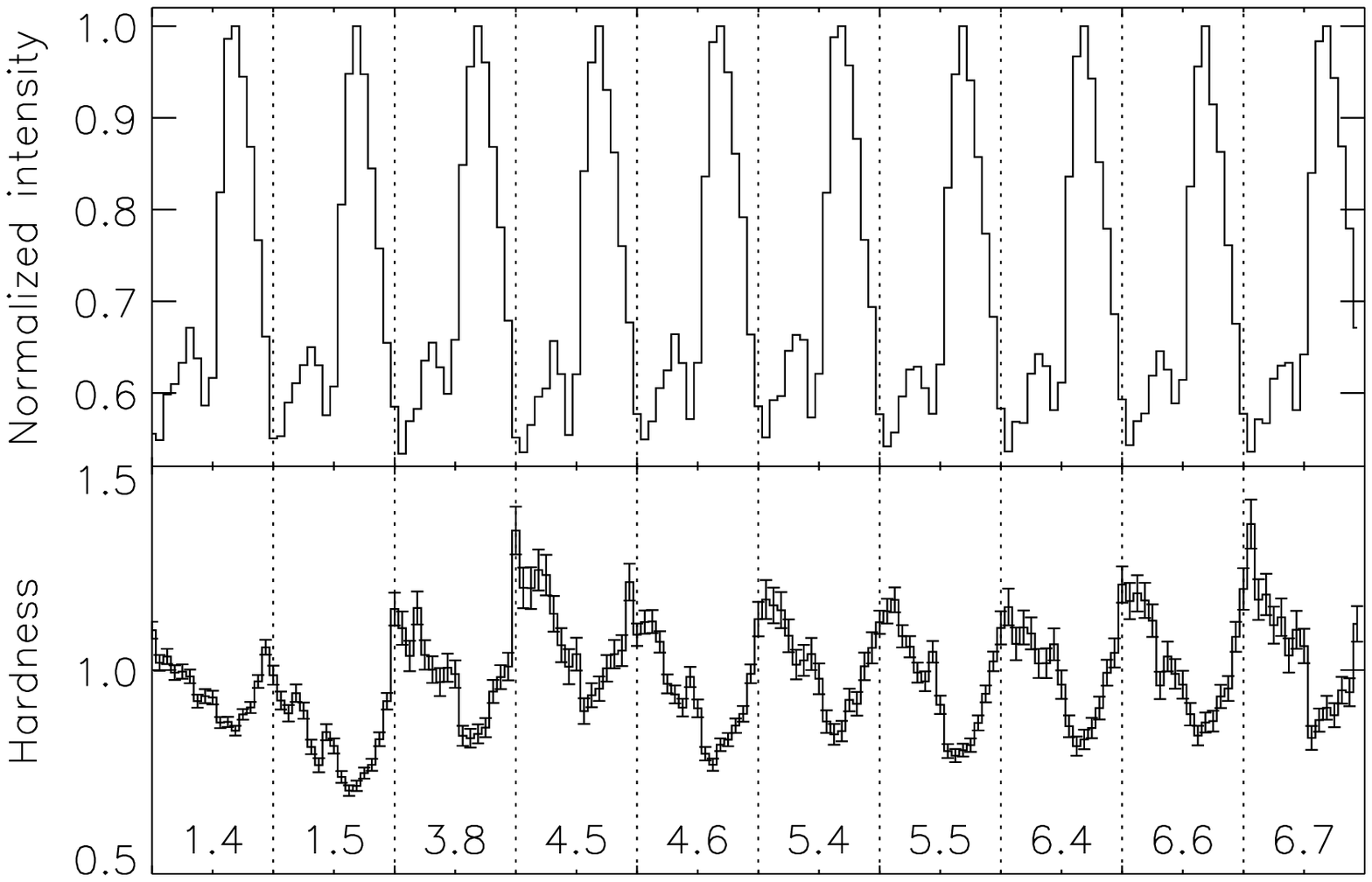} \\
\vspace{-0.8cm}
\includegraphics[width=8cm]{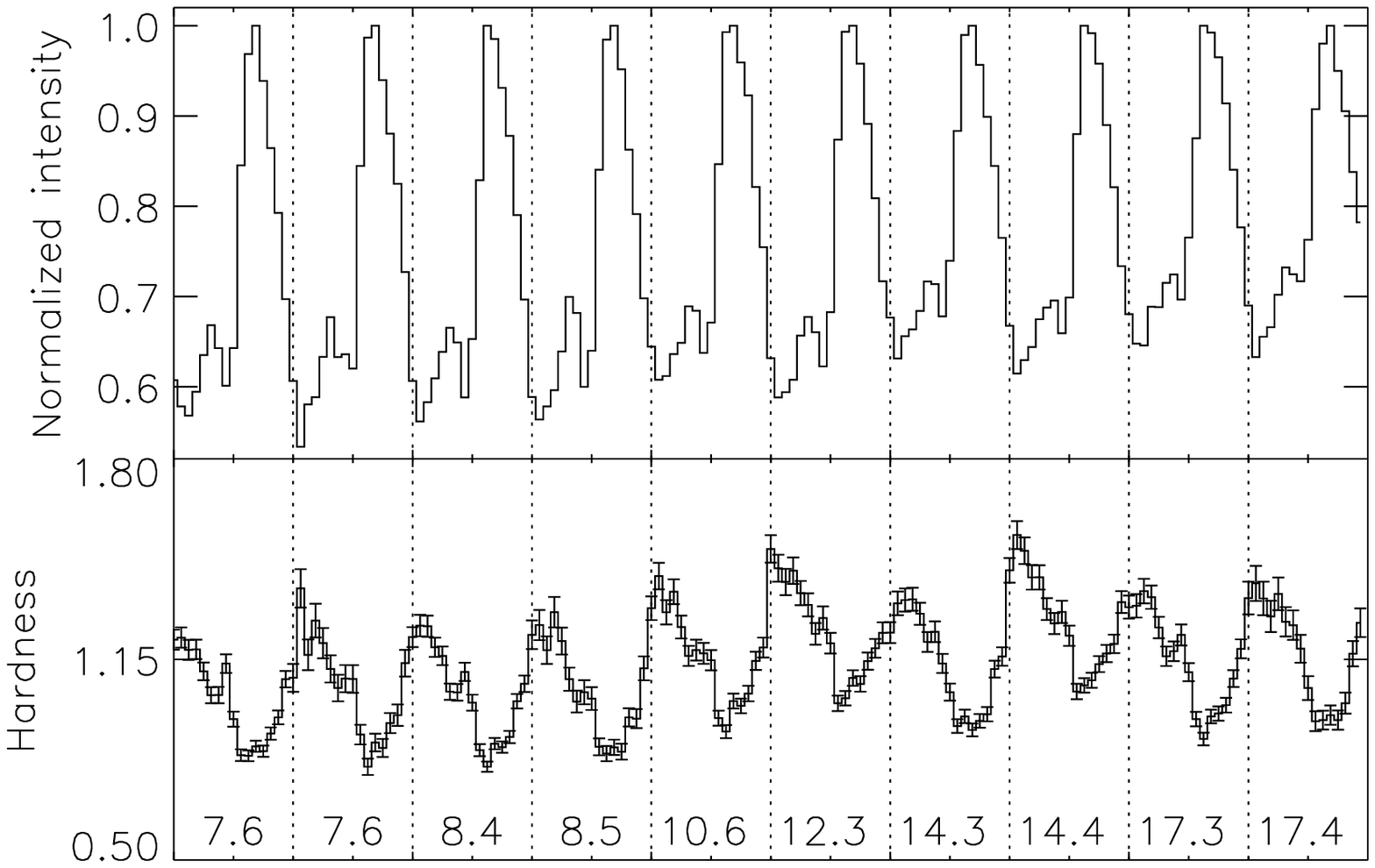} \\
\vspace{-0.8cm}
\includegraphics[width=8cm]{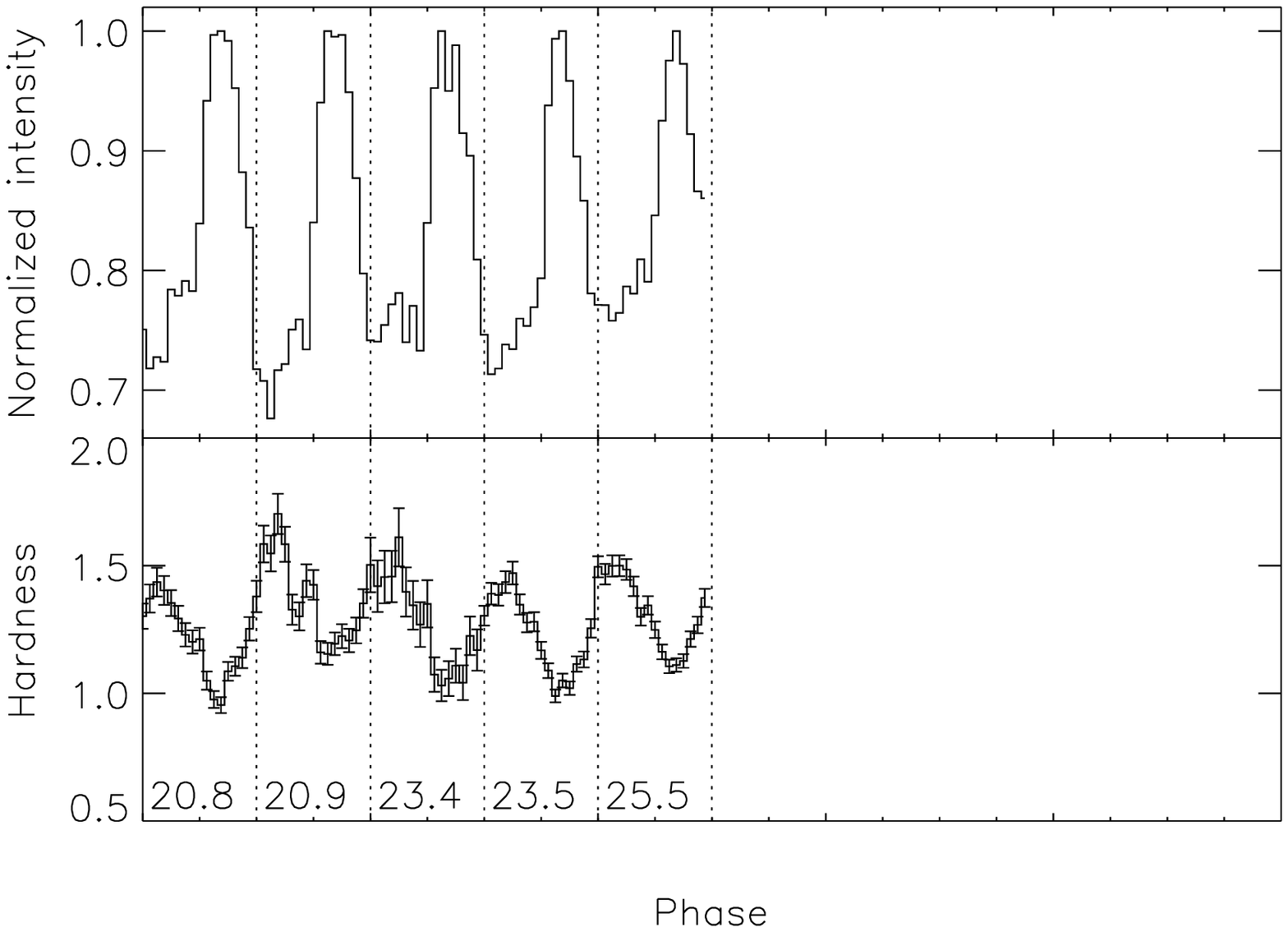} \\
\end{tabular}
\vspace{0.8cm}
\caption{Evolution of the pulse profile and its hardness ratio for the first 25 days of the outburst. In each panel, the upper curve is the pulse profile and the lower curve is the hardness ratio. The vertical dotted lines separate individual observations; their times relative to the outburst onset (2011 July 15) are indicated at the bottom of the plot. \label{hrprof}}
\end{figure}	

We further studied this matter by looking at the HR of the pulse profiles constructed from each observation during the outburst. To create the pulse profiles, we used the spin ephemeris provided by \citet{Rea2012}: spin frequency 0.118515426(3) Hz and its time derivative $1.17(3) \times 10^{-15} s^{-2}$ at the epoch 55757 MJD. We generated pulse profiles in the two energy bands: 5.0--14.0 and 2.0--5.0 keV and calculated the HR by taking the ratio of the pulse profile intensities in these two bands. The HR of pulse profiles was then plotted along with the pulse profile for each observation during the first 25 days of the outburst. In Figure~\ref{hrprof} we present the variation of the normalized pulse profile intensity in the 2.0--14.0 keV energy band and the corresponding HR during the outburst. We find that the HR shows strong anti-correlation with the pulse profile intensity, as was already prominent from our phase lag analysis. To quantify this, the rank correlation coefficient between the folded profile and its hardness was computed and its value was obtained to range from $-0.691$ (with a chance probability of $3.02\times10^{-3}$) to $-0.956$ (with a chance probability of $7.76\times10^{-9}$).

\subsection{Pulse profile analysis}\label{pulseprof}

\begin{figure}[H!ba]
\centering
\includegraphics[width=8cm]{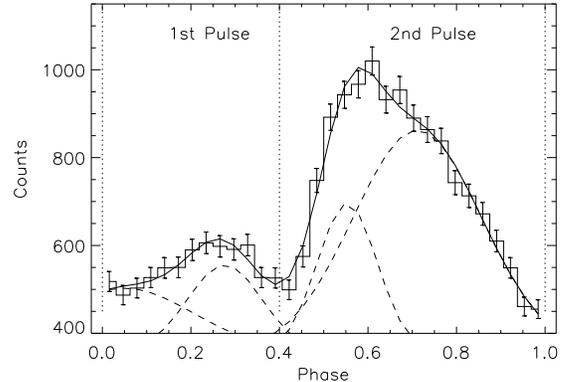}
\caption{Example pulse profile obtained with PCA on 2011 July 19 to illustrate the pulse complexity. The pulse profile contains sub-structures which could be well-fitted using a combination of 4 Gaussian components shown in the plot with dashed lines. The broad sub-pulses mentioned in the text are pointed out, and seperated by the vertical dotted line. \label{profile}}
\end{figure}

We also investigated the morphological variations of the pulse profile of Swift J1822.3--1606 both in time and with energy through the early phases of its 2011 outburst. To perform this analysis, we again used the spin ephemeris obtained by \citet{Rea2012} as employed in the earlier sub-section. We folded the data collected with all operational PCUs to get the phase folded light curves for each observation day. To form the pulse profiles, we used 16 or 32 phase bins, depending on the signal strength. Pulse profiles of Swift J1822.3--1606 were complex initially following the outburst onset, containing various sub-structures or sub-pulses.
In Figure~\ref{profile} we show the pulse profile of the source on 2011 July 19, that is 5 days into the outburst, to introduce the sub-structures in the pulse profile. 

\begin{figure}[H!ba]
\centering
\begin{tabular}{c}
\includegraphics[width=8cm]{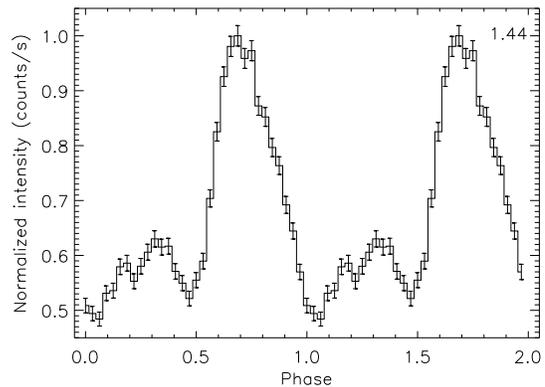} \\
\includegraphics[width=8cm]{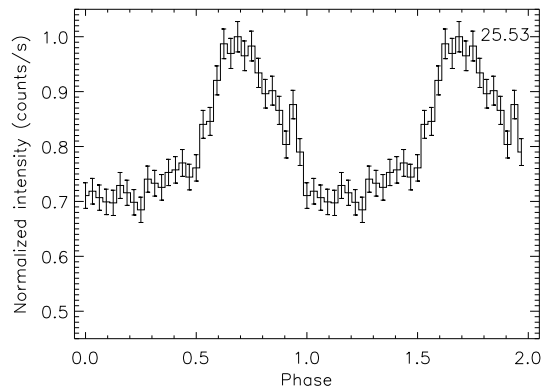}
\end{tabular}
\caption{Evolution of the $2-10$ keV pulse profile at the early (top panel) and later (bottom panel) stages of the 2011 outburst of Swift J1822.3--1606. The corresponding time of each profile is given in days since 2011 July 15 at the top right corner of each panel. \label{profevol}}
\end{figure}

To investigate the profile in more detail, we initially fit the sub-structures using a model comprised of Gaussian components. During the early stages of the outburst, a combination of 4 Gaussian functions was required to describe the profile well as displayed with the dashed lines in Figure~\ref{profile}. This is consistent with the observations of \citet{Feroci2001} and \citet{Mazets1999} who detected clear 4-peaked sub-structures in the pulse profiles observed during the decaying tails of the giant flare from SGR 1900+14. As the outburst decays, the sub-pulses become less prominent, therefore, the overall profile becomes less complex and exhibits a nearly sinusoidal like shape. In Figure~\ref{profevol}, we present the $2-10$ keV pulse profiles at the onset and after 25 days during the outburst. It is clearly observed that the complexities of the pulse profile diminish as the outburst decays. Similar pulse profile evolutions from complex to simpler morphology have been previously reported following the giant flare of SGR 1900+14 \citep{Mazets1999, Feroci2001, Woods2001}.  

We found that the phases of the sub-pulses (that is the centroids of the four Gaussians) remained constant within errors during the early episodes of the outburst. Among the four Gaussian components, the first two and the last two sub-structures could not be clearly resolved from each other. Moreover, it becomes difficult to constrain the model parameters of the four Gaussian components as the X-ray intensity gets weaker. For these reasons, we considered only the two clearly resolved broad sub-pulses (see Figure~\ref{profile}) and estimated the contributions of these sub-structures to the pulse profile using the following method. We identify the two sub-structures as ranging between phases $0.0-0.4$ and $0.4-1.0$ respectively, which are significantly detectable for the entire outburst. The two broad components of the pulse profile are indicated by dotted vertical lines in Figure~\ref{profile}. 
In order to quantify the contributions of the sub-structures to the pulse profile, we calculated the rms pulsed count rate (PCR) for the sub-pulses. We used the definition of the PCR
\begin{equation}
PCR=( \frac{1}{N} \sum_i^N ((R_i-R_{avg})^2 - \Delta R_{i}^2) )^{\frac{1}{2}}
\end{equation}
where $R_i$ is the count rate, $\Delta R_{i}$ is the uncertainty on the count rate in the $i$th phase bin and $N$ is the number of phase bins within each sub-pulse, and $R_{\rm avg}$ in the average count rate in the entire pulse profile. Note that the second term in Equation 1 ($\Delta R_{i}^2$) accounts for any possible bias in the rms PCR due to statistical uncertainties, and is routinely employed in the literature (see, e.g., \citealt{Gogus2002, Dib2008, Dhillon2009, Gonzalez2010, Gavriil2011}). We calculated the PCR values for each sub-pulse for each observation using data from PCU2 since it was the only unit operational in our entire data set. This way we could obtain a uniform view of the pulse profile evolution without having any possible cross-calibration issues of mixed PCUs. The PCRs were calculated in the $2-10$ keV energy range for both sub-pulses.

\begin{figure}[H!tba]
\centering
\begin{tabular}{c}
\includegraphics[width=8cm]{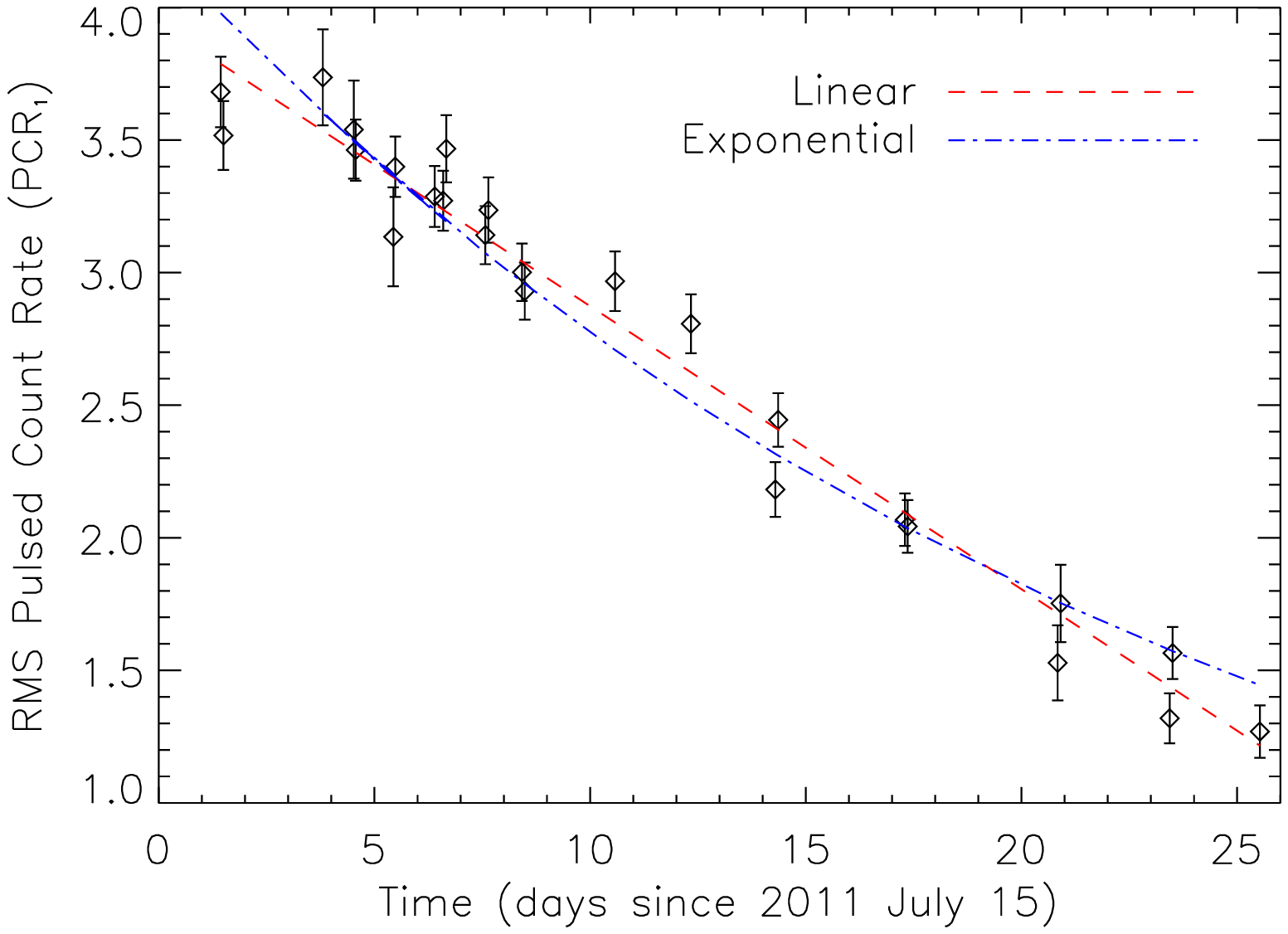} \\
\includegraphics[width=8cm]{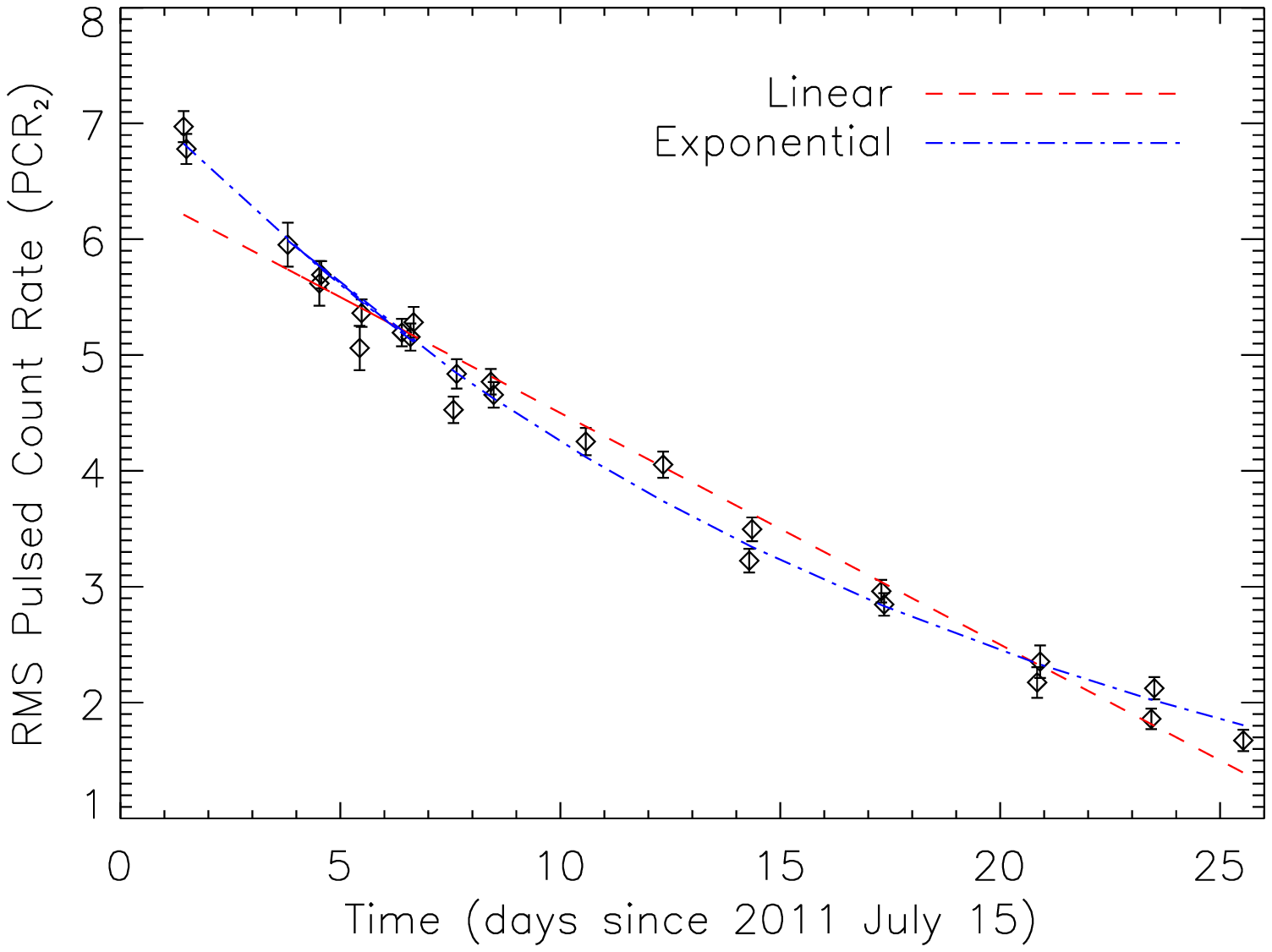}
\end{tabular}
\caption{Variation of rms pulsed count rate (PCR) with time for the two sub-pulses. The top panel shows the temporal evolution of PCR of the first sub-pulse and the bottom panel shows the same for the second sub-pulse. The fits to evolution using linear and exponential model are shown by red dashed and blue dot-dashed lines, respectively. \label{PCRtime}}
\end{figure}

We find that the PCR values of both pulse components decreased steadily as the outburst progressed, which is in accordance with what was reported in \citet{Livingstone2011}. We modeled the temporal evolution of the PCR of each sub-pulse using two functions: a linear and an exponential model. Figure~\ref{PCRtime} shows the PCR variation of each of the sub-pulses during the outburst, along with the best fit model curves. For the time evolution of the first sub-pulse, the two models yield reduced $\chi^2$s of 1.3 and 2.5 for 23 and 23 degrees of freedom, respectively. For the second sub-pulse, $\chi^2_{\nu}/\nu$ were obtained as 2.7/23 and 0.9/23 for the linear and exponential models, respectively.

\begin{figure}[H!tba]
\centering
\includegraphics[width=8cm]{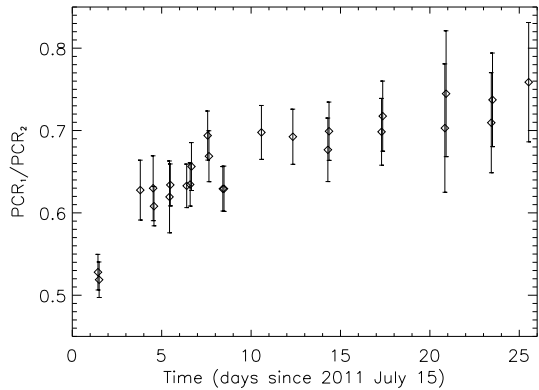}
\caption{The variation of the ratio of the rms pulsed count rate (PCR) of the first sub-pulse (PCR$_1$) to that of the second sub-pulse (PCR$_2$) with time. The ratio shows a marginal increase trend with time, implying that the contribution of the weaker sub-pulse to the pulse profile increases with time compared to the stronger one. \label{PCRratio}}
\end{figure}

We observed that the first and the second sub-pulses evolved differently in time during the early phase of the outburst as the temporal variation of the former pulse component was well described by a linear model whereas the temporal variation of the latter followed an exponential trend. The decay constant of the exponential fit was obtained to be $23.8\pm0.8$ and $18.1\pm0.3$ for the first and second sub-pulses, respectively, which are close to the decay constant obtained by \citet{Livingstone2011} for the temporal evolution of the entire pulse. To check the relative contributions of the two sub-pulses, we determined the ratios of the PCR trend of the first sub-pulse to that of the second one, and present the evolution of the ratios in Figure~\ref{PCRratio}. We found that the contribution of the first sub-pulse to the pulsed emission was low initially ($\lesssim30\%$), but increased in a time frame of days to about $45\%$.

\begin{figure}[H!tba]
\centering
\includegraphics[width=8cm]{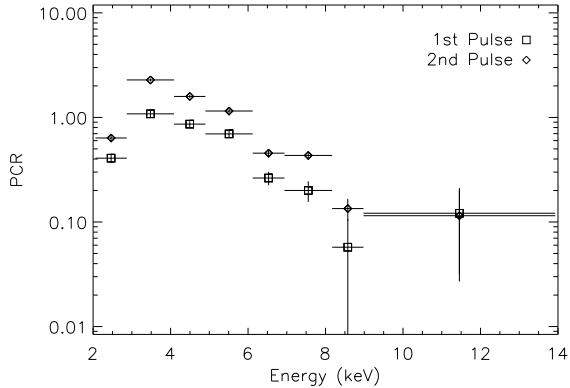}
\caption{Typical example of the variation of rms pulsed count rate (PCR) with energy using PCA observation performed on 2011 July 16. The first and the second sub-pulse are denoted using square and diamond symbols, respectively. \label{PCRen}}
\end{figure}

Motivated by the fact that the pulse amplitude was stronger at the softer energy bands and gets weaker as the energy increases, we also examined the energy dependence of of each pulse component. To quantify this, we computed the PCRs of each of the two sub-pulses considering different energy ranges,-- $2.06-2.87$, $2.87-4.09$, $4.09-4.90$, $4.90-6.12$, $6.12-6.94$, $6.94-8.17$, $8.17-8.98$, and $8.98-14.0$ keV (see Figure~\ref{PCRen}). This can be regarded as a crude representation of the pulsed X-ray emission spectrum of the two pulse components. We present the variation of the PCRs of the two sub-pulses with energy ranging from 2-14 keV for the observation on 2011 July 16, in Figure~\ref{PCRen}. It is evident that both pulse components varied similarly in energy: at first they both increased slightly, then reached a peak around 3.5 keV, and then decreased steadily with increasing energy. We found that the ratio of the PCR of the first sub-pulse to that of the second sub-pulse remained constant in energy considering errors for each observation. The values of the ratio for 2011 July 16 observation varied from $0.427\pm0.379$ to $1.056\pm1.123$ with a mean of 0.598 and a standard deviation of 0.199.

\begin{figure}
\centering
\includegraphics[width=8cm]{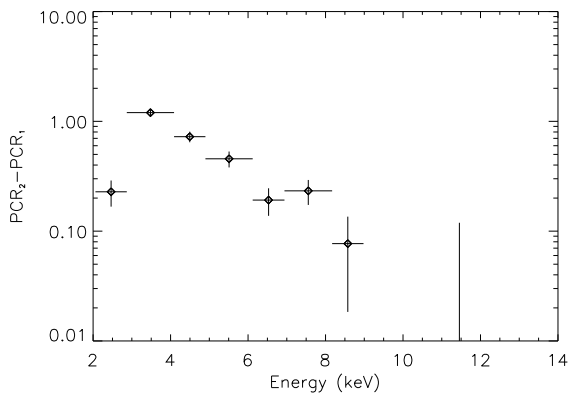}
\caption{Example of the behavior of the difference between the rms pulsed count rate of first sub-pulse (PCR$_1$) and that of the second (PCR$_2$) with energy for the same observation employed in Figure~\ref{PCRen}. \label{PCRend2}}
\end{figure}

We also find that at the high energies, the pulses became weaker and the contributions of the two pulse components to the pulsed emission became comparable. This is clearly illustrated in Figure~\ref{PCRend2} where the difference in PCR values between the two sub-pulses is plotted as a function of energy. The difference between the PCR values of the two sub-pulses followed a behavior much like PCR values themselves, increasing initially and then decreasing steadily with energy. The ratio of the of values of the PCR of the two sub-pulses did not show much variation with energy as the difference is quite small and follows a similar behavior to the individual PCR values.

\subsection{Search for quasi-periodic oscillations}

Quasi-periodic oscillations were observed during the tails of giant flares observed from SGR 1806--20 and SGR 1900+14 \citep{Israel2005, Strohmayer2005} in the frequency range from tens of Hz to few 100 Hz. Therefore, we also checked whether such QPO-like timing features are present in power spectra of the tail of the outburst of Swift J1822.3--1606. We created Leahy-normalized power spectra \citep{Leahy1983} from the filtered light curves in the energy range of 2.0-40.0 keV. To detect significant signals, we binned the power spectrum employing a number of techniques as described next and then searched for peaks in the resultant binned power spectrum. First, we determined the power spectrum of every 1 s of the time series and averaged them to obtain a power spectrum with reduced noise contribution. For this task, we considered a Nyquist frequency of 2048 Hz. For each averaged power spectrum, we searched the power spectrum for signal power excluding any red noise component, considering all powers above 1 Hz. We calculated the single trial significance for the maximum detected power and multiplied it by the number of trials which is the number of frequency bins searched. In the end, we did not detect any clear signal with a significance greater than $3 \sigma$ i.e., where the probability of the signal originating purely due to noise is $2.7\times10^{-3}$ or less. We then performed a search of each 1 s power spectra in the frequency range $1-1020$ Hz, using a Nyquist frequency of 1024 Hz, in order to account for the case of quasi-periodic signal appearing only for a very short time interval. Our search yielded no signal more significant than $3 \sigma$ considering the number of trials i.e, the number of frequency bins searched times the number of power spectra searched. 

We then proceeded to check whether there exists any energy dependent timing features in the power spectra of this source. For this purpose,  we repeatedly performed the above procedure for four energy ranges: $2.0-4.5$, $4.5-8.5$, $8.5-14.0$ and $14.0-40.0$ keV.  The resulting power spectra did not reveal any significant timing features in any of the energy ranges.

Finally, we combined the two procedures described above. We considered relatively small time segments of width 10 s, extracted power spectra in each 1 s within these time intervals and averaged over them to get a resultant power spectrum for each 10 s segment. This ensures that if the signal is present for a small interval of time, it will not be smoothened by adding a large number of power spectra where the signal is absent and moreover the averaging over the interval reduces the noise contribution lessening the chance of false detection. We performed this analysis also over the above-mentioned 4 energy channels.  From our searches, we did not obtain any significant detection greater than $3 \sigma$ considering all the number of trials.

\section{Discussion}\label{Discussion}

We have performed a detailed investigation of temporal variations of the X-ray intensity and HR of Swift J1822.3--1606 throughout the early phases of its 2011 outburst episode. We found the following interesting properties, which are quite similar to the behavior observed in the decaying tail of the magnetar giant flares, in particular, the 1998 August 27 giant flare from SGR 1900+14.
\begin{enumerate}
\item Swift J1822.3--1606 exhibited clear periodic modulations at the spin period of the neutron star during the onset of its outburst, which is quite alike the decaying tail of the giant flares.  
\item There is a strong anti-correlation between the HR and the intensity, specifically the pulsed intensity. The same behavior was also observed in the tail of SGR 1900+14 giant flare. The HR for both cases show ramp or see--saw like patterns with pulse profile intensity. The pulse peaks corresponded to the dips in the HR for both cases. For the 2011 outburst decay of Swift J1822.3--1606, we unveiled that the hardness lags behind the pulse intensity and the value of the phase lag is consistent with the observed anti-correlation.
Our results strengthen the idea of intensity-hardness anti-correlation due to Comptonization \citep{Thompson2001}, since radiation has to propagate a much longer physical length of magnetospheric medium when the trapped fireball is away from the observer (pulse minimum). Therefore, emerging radiation during pulse minimum undergoes more Compton upscatterings.
\item The pulse profiles in both cases exhibit complex structures, in particular a 4-peaked structure. For the case of the giant flares, the complex pulse structure gets smoothened to an almost sinusoidal shape within minutes. Much like the giant flares, the initially complex pulse profile of Swift J1822.3--1606 became much less complicated, but on a longer timescale of about $20-30$ days.
\item The phases of the sub-pulses are quite stable with respect to each other and also with time for both the giant flare of SGR 1900+14 and the 2011 outburst of Swift J1822.3--1606.
\item The decay of the persistent X-ray flux in both cases are well described by a double exponential function.    
\end{enumerate}

Based on these similarities, we suggest that the same underlying mechanism is likely taking part in triggering the giant flares and the much longer lasting outbursts of magnetars. In both cases, the triggering mechanism is possibly a profound crustal deformation \citep{Thompson1995, Thompson2001} or a magnetospheric activity \citep{Lyutikov2003, Beloborodov2013, Lyutikov2015}. In the first scheme, what differs the giant flares from the longer lasting outbursts is not the way the energy is injected, but how the injected energy is radiated away. Namely, we suggest that the giant flare and the 2011 outburst of Swift J1822.3--1606 are observational manifestations of the same phenomena, while the latter is a somehow suppressed form of the former, so that a similar amount of injected energy is radiated away less efficiently in a much longer time frame.

In both crustal deformation or magnetospheric activity schemes, a ``trapped fireball'' of energetic pairs and photons forms and is accounted for the decaying tail of the giant flares. According to the second scheme, no crustal fracture is needed and the flares and bursts originate from the magnetospheric reconnection events influenced by the electron magnetohydrodynamic currents \citep{Lyutikov2015}. In this picture, the giant flares can occur arising from a large-scale restructuring of the magnetic fields only if the magnetic field of the neutron star is sufficiently strong. For systems with weaker magnetic fields, the outcome is expected to be typical short bursts and enhanced persistent emission due to the magnetic field evolution \citep{Lyutikov2015}. Our results are in agreement with this prediction since Swift J1822.3--1606 has a relatively low dipole magnetic field as inferred from its spin behavior.

For the 1998 August 27 giant flare of SGR 1900+14, the energy released in the tail in the energy range of 40-700 keV was estimated as $\sim 10^{42}$ erg \citep{Feroci1999}. To compare with what is observed during the decay of Swift J1822.3--1606 outburst, we calculated the emitted flux by performing spectral analysis of a subset of contemporaneous \textit{Swift} XRT observations. We found that the 0.5-10 keV flux was $1.96\times10^{-10}$ erg s$^{-1}$ cm$^{-2}$ on July 17, i.e., soon after the onset of the outburst, then it was $\sim10^{-10}$ erg s$^{-1}$ cm$^{-2}$ on July 27, and $0.6\times10^{-10}$ erg s$^{-1}$ cm$^{-2}$ on August 8. Assuming the distance as $1.6$ kpc \citep{Scholz2012}, these flux values yield luminosities of $6\times 10^{34}$ erg s$^{-1}$, $3.2\times 10^{34}$ erg s$^{-1}$, and $1.8\times 10^{34}$ erg s$^{-1}$, respectively. We then calibrated the \textit{Swift} XRT with these flux measurements and calculated the X-ray flux over 25  days following the BAT trigger. We modeled the evolution of the outburst flux decay using a single exponential function, with a decay constant of 10.8 days. Finally we integrated the flux over 25 days from the BAT trigger to obtain the total fluence, and thus the total emitted energy in the band pass of 0.5-10 keV was found to be $8.4 \times 10^{40}$ erg.
We would like to point out that this value is only a lower limit on the total energy released in the outburst. We also looked at the \textit{Swift} BAT data in order to check the contribution of the high energy emission to the whole outburst energetics. For BAT analysis, we employed survey mode data, and used a HEASOFT routine {\tt batsurvey} which runs various filtering, re-binning and cleaning procedures described in detail by \citet{Tueller2010}. We obtained the $15-150$ keV flux on 2011 July 15 (i.e., the outburst onset) as $2 \times 10^{-10}$ erg s$^{-1}$ cm$^{-2}$ which corresponded to a luminosity of $6.2\times10^{34}$ erg. BAT observations in the following days did not reveal any significant detection of emission at the position of the source, implying that the high energy emission of Swift J1822.3--1606 faded in less than a day. Such a short duration for hard X-ray emission is likely related to the life time of energetic particle loaded corona, which could up-scatter emitted soft X-ray photons \citep{Baring2007}.

To provide a conservative limit on the timescale required to release this energy, we divided the energy released in the giant flare by the average flux of the outburst decay. We found a timescale of about 200 days. Note that the temporal evolution of the flux was modeled with a double exponential model for both cases: the time constants were 15.5 days and 177 days for the outburst and 5 s and 80 s for the giant flare. From Figure 8 of \citet{Scholz2012} it can be seen that for a injection energy of $10^{42}$ erg, the source requires about a few hundred days to release the deposited energy considering the observed flux values. \citet{Rea2012} also showed similar evolution of the outburst intensity and using the theoretical model put forward by \citet{Pons2012}, they modeled the outburst decay. 

\citet{Livingstone2011} showed that the entire pulsed flux evolution of the outburst follows a single exponential model. However, we found that the pulse profile is highly complex and contains sub-structures. The time evolution of these sub-structures carries invaluable information about the variations of the neutron star surface emission topology. Here, we divided the pulse profile into two broad sub-structures (see Figure ~\ref{profile}) and found out that these two components behave in different manners. Though both decrease with time, the evolution of the first sub-pulse is better explained by a linear decay whereas the evolution of the second sub-pulse follows an exponential trend. This implies a slightly different radiative behavior for the two. Moreover, the first sub-structure is weak initially compared to the second one, while it becomes relatively stronger as the outburst progresses, and reaches about 75\% of the intensity of the second sub-structure at the end of 25 day time segment.
We also presented that the pulsed flux of both pulse components peaked around 3.5 keV, and both declined monotonically with energy above 4 keV. Therefore, both sub-components of the pulse evolving in time and energy in similar manner, likely experienced similar radiative behavior, yielding two major trapped fireballs emitting throughout the outburst decay.

\section*{Acknowledgments}
M.C. is supported by the Scientific and Technological Research Council of Turkey (T\"{U}B\.{I}TAK) through Research Fellowship Programme for International Researchers (2216). We thank Amy Y. Lien for valuable suggestions regarding analysis of \textit{Swift} BAT survey data.

{}

\end{document}